\documentclass[12pt,preprint]{aastex}

\newcommand\Mpc{$h^{-1}$Mpc}
\newcommand\etal{{\it et al.\ }}
\newcommand\kmsec{${\rm km~sec^{-1}}$}
\newcommand\sig{\sigma_{12}}

\begin{document}

\title{The Pairwise Velocity Distribution Function of Galaxies in the LCRS,
                2dF, and SDSS Redshift Surveys}

\author{Stephen D. Landy}

\affil{Department of Physics, College of William and Mary,
     Williamsburg, VA 23187-8795
        landy@physics.wm.edu}

\begin{abstract}

    A comparison of the galaxy pairwise velocity distribution
functions determined from the three largest publicly available galaxy redshift
surveys is presented. This is the first direct comparison of this function
across these surveys using an identical method. It is found that the two r-band
selected surveys, the LCRS (Las Campanas Redshift Survey) and the SDSS (Sloan
Digital Sky Survey), are in excellent agreement with $363 \pm 44$~\kmsec~ and
$357 \pm 17$~\kmsec~ respectively. The b-band selected 2dF survey gives
slightly lower results with $331 \pm 19$~\kmsec. This difference is expected
given the sampling biases in the different surveys although it is not highly
significant. These results and analysis of subsets of data indicate that the
method is stable to singular features in each survey such as the prevalence of
rich clusters. The technique utilizes a Fourier-space deconvolution of the
redshift-space distortions in the correlation function in the $(r_{p},\pi)$
basis, which has been previously described. This method returns the entire
distribution function rather than just the second moment. In all cases the
distribution function is well-characterized by an exponential.

\end{abstract}


\section{Introduction}

If, as generally accepted, clustering evolves through gravitational instability
then the properties of galaxy clustering and the distribution of galaxy
peculiar velocities contain fundamental information on the global properties of
the Universe and provide important constraints on cosmological models. For a
given initial mass fluctuation spectrum, clustering evolution and the
generation of galaxy peculiar velocities is strongly dependent on the value of
the mass density parameter $\Omega_m$. For example, in a low $\Omega_m$
Universe mostly devoid of matter, it is very difficult to generate high
peculiar velocities since the Hubble flow will dominate over gravitational
instability from early times unless the mass fluctuation amplitude is very
high. Consequently, low values of $\Omega_m$ generate relatively smaller
peculiar velocities than high values for a given fluctuation amplitude. Once
both the present galaxy clustering amplitude and peculiar velocity distribution
functions are well-determined, cosmologists will have made significant progress
in determining the initial conditions and evolution of the mass density in the
Universe.

Unfortunately, neither galaxy positions nor peculiar velocities are presently
measurable with significant accuracy to make robust, direct measurements of
these two quantities. Rather, most of the information available today is
contained in galaxy redshift surveys. As is well-known, the redshift to a
galaxy represents the sum of the galaxy's `distance' and its radial peculiar
velocity. Therefore, measurements of the galaxy two-point correlation function
are contaminated by these redshift distortions and what is measured is
appropriately called the redshift-space correlation function (see also
Juszkiewicz \etal 2000).

It is possible, however, to exploit the anisotropies created by the peculiar
velocities in the redshift-space correlation function (see Peebles 1980). In
this method, the redshift space correlation function is represented in
two-dimensions, where the axes correspond to the directions parallel ($\pi$)
and perpendicular ($r_{p}$) to the line-of-sight. The resultant correlation
function $\xi_{z}(r_{p},\pi)$ is then anisotropic since the the peculiar
velocities distort the correlation function principally along the
line-of-sight. Since the galaxy-galaxy correlation function is a two-point
estimator, the anisotropies being generated actually reflect the value of the
{\it pairwise} galaxy peculiar velocity distribution function. Measurement of
this distribution function is the focus of this {\it Letter}.

In a seminal paper, Davis and Peebles (1983) using redshift information from
the first Center for Astrophysics Redshift Survey (CfA1) measured a value for
the second moment of the pairwise velocity distribution function (hereafter
PVD) of $\sigma_{12} = 340\pm40$~\kmsec~ and characterized the distribution as
an exponential. Subsequent numerical work predicted a much larger value of
approximately 1000~\kmsec~for a standard $\Omega h=0.5$ Cold Dark Matter (CDM)
model (Davis \etal 1985). However, later work has questioned the accuracy of
the original CfA1 result (Mo, Jing, \& B\"{o}rner 1993, Somerville, Davis \&
Primack 1997).

Subsequent redshift surveys have given somewhat discrepant results. The IRAS
1.2 Jy survey (Fisher \etal 1994) was in good agreement with a value of $\sig=
317^{+40}_{-49}$~\kmsec~while Marzke \etal (1995) using the second Center for
Astrophysics Redshift Survey (CfA2) together with the Southern Sky Redshift
Survey (SSRS) found $\sig= 540\pm180$~\kmsec. Similar analysis applied to the
Las Campanas Redshift Survey, hereafter LCRS, has found $\sig=452\pm60$~\kmsec
(Lin \etal 1995). Guzzo \etal (1997) find $\sig= 345^{+95}_{-65}$~\kmsec~for
late-type galaxies using the Pisces-Perseus Redshift Survey and report this to
be a fair estimate for field galaxies. Other recent measurements include Small
\etal (1999) using the Two Norris Redshift Surveys who reports
$326^{+67}_{-52}$~\kmsec~ and Ratcliffe \etal (1998) using the Durham/UKST
Galaxy Redshift Survey with an estimate of $416 \pm 36$~\kmsec.

Numerical work and re-analysis of existing surveys have shown the sensitivity
of the measurement of the pairwise velocity dispersion to the idiosyncracies of
the data using standard techniques. Mo \etal (1993) found that the estimated
dispersion is extremely sensitive to the presence of rich clusters in a sample,
$\sig=$ 300 to 1000 \kmsec~ for subsets of the same data. Zurek \etal (1994)
using high resolution CDM simulations found large variations in the value of
the dispersion on CfA1 size scales. Using real data, Marzke \etal (1995) found
that by excluding the rich clusters from their survey, R$\ge 1$, the measured
velocity dispersion dropped to $\sig=295\pm99$~\kmsec.

The sensitivity of the standard methods to the presence of rich clusters is
primarily due to the fact that they generally only estimate the second moment
of the distribution, which is highly sensitive to the existence of hot, rich
clusters in the data. Recognizing this problem, several authors have invented
new statistics that are less sensitive to rich clusters (see Kepner, Summers \&
Strauss 1997, Davis, Miller \& White 1997, Strauss, Ostriker \& Cen 1998, and
Baker, Davis \& Lin 2000). While new statistics can be a powerful approach,
their measures are often somewhat indirect and their results less intuitive.

To circumvent this problem and with the intent to determine the entire peculiar
velocity distribution function rather than just the second moment, Landy,
Szalay \& Broadhurst (1998), hereafter LSB98, developed a method based on a
Fourier-space deconvolution of the redshift-space distortions in the
correlation function in the $(r_{p},\pi)$ basis. Using this approach with the
LCRS data, a value of $363 \pm 44$~\kmsec~ was obtained. Further, by recovering
the Fourier transform of entire distribution function, it was directly shown
that the PVD is well-characterized by an exponential.

Other recent measurements of the PVD using the same LCRS data but different
techniques have also been reported. Jing, Mo, \& B\"{o}rner (1998) measured the
PVD of the LCRS data and obtained $570 \pm 80$~\kmsec. The discrepancy with the
result of LSB98 was ascribed to a failure of LSB98 to account for infall
effects (see Jing \& B\"{o}rner 1998). Another analysis of the LCRS data was
presented in Jing \& B\"{o}rner (2001), using a method closely related to that
of LSB98, with a PVD value of $510 \pm 70$~\kmsec. The discrepancies between
these two latter results and those in this {\it Letter} will be presented in
the discussion.

Recently, Peacock \etal (2001) have published a measurement of the PVD using a
larger sample of the 2dF survey than publicly released as an adjunct to their
estimation of the cosmological parameter $\beta$. They obtain a value of
$385$~\kmsec~ but do not report standard errors.

\section{Data}

The three data sets used in this analysis are all publicly available. The LCRS
data is described in Shectman \etal 1996. The results for that data published
here are identical to those reported in LSB98 and the reader is directed there
for further information.

The 2dF data is the release of June 2001 (see www.mso.anu.edu.au/2dFGRS/). This
data has been cut to $0.05 < z < 0.2$ (150 to 570 \Mpc~assuming $\Omega_m =
0.3, \Omega_{\lambda} = 0.7$) and is naturally broken up into two subsets
toward the north and south galactic caps. These sets are disjoint on the scales
under consideration. The north and south data consisted of 34805 and 42309
galaxies respectively. Random catalogs for this data were generated using the
publicly available routines supplied by the 2dF Survey.

The SDSS data is the release of June 2001 (see Stoughton \etal 2001). As with
the 2dF data, the redshift limits were set at $0.05 < z < 0.2$ and subsets of
the data towards the north and south galactic caps were analyzed independently.
This data consisted of 8918 galaxy positions in the North and 7862 in the
South. The random catalogs for this data were kindly supplied by Adrian Pope.

\section{Method}

In essence, this approach extracts the Fourier transform of the PVD from the
Fourier transform of the galaxy-galaxy correlation function in the
$(r_{p},\pi)$ basis using a deconvolution procedure. By working in Fourier
space, the deconvolution is one of simple division that directly returns the
Fourier transform of the PVD. Since the method here is identical to that
described in LSB98, the reader is directed there for further details. However,
given the stability of the reported results, it is worth re-iterating two of
the principle advantages of working in Fourier space.

Firstly, by utilizing the Fourier transform of the $\xi(r_p,\pi)$ correlation
function, the measurement of the PVD is no longer a pair-weighted statistic.
This reduces the contamination by rich clusters in the data. More specifically,
the presence of rich clusters in the sample super-imposes a number of pairs
with a higher dispersion than the thermal dispersion of field galaxies. These
pairs predominately effect the tail of the correlation function and produce an
overestimation of the PVD, especially of its second moment. By determining the
Fourier transform of the PVD rather than the PVD itself, the large number of
bins in the tail are all compressed into the innermost few resolution elements
in Fourier space. Therefore they are naturally down-weighted in the fitting
procedure, which uses equal weights for every cell in $k$-space. This method
does not exclude the signal from the rich clusters but rather incorporates them
in a way which decreases their contribution to the variance in a robust
fashion.

Secondly, in the deconvolution procedure, the weighting function is effectively
proportional to $r_p^{-1}$, which emphasizes the high signal-to-noise core of
the correlation function. Since most of the mass of the correlation function
lies within the central core, the resulting signal will predominately reflect
the value of the PVD within a scale of about 1\Mpc. This both reduces
contamination by infall effects, which are expected to be small on these
scales, and localizes the measurement to reflect the value of the PVD on small
scales.

\section{Results}

The Fourier transform of the PVD for each redshift survey is shown in Figure 1.
In all cases, this function is well-characterized by a Lorentzian. Since the
Fourier transform of a Lorentzian is an exponential, the PVD itself is
well-characterized by an exponential. For comparison, the best-fitting Gaussian
distributions are also shown. All values reported correspond to the equivalent
exponential value for a given fitted Lorentzian.

The result is very stable across all three redshift surveys. Since the 2dF is a
b-band selected survey and the LCRS and SDSS are both r-band selected, it is
expected that the 2dF Survey will contain a relatively greater number of field
galaxies. Thus, the PVD for the 2dF survey should have a lower value as is
found: $357 \pm 17$~\kmsec~for the SDSS versus $331 \pm 19$~\kmsec~for the 2dF.
However, this difference is not of high significance given the standard errors
for the data.

The Lorentzian in the case of the 2dF data shows additional structure around
the central peak. This type of structure has also shown up in some of the
analyses of subsets of the data. The effect is likely a consequence of the fact
that the method necessarily constrains the value of the signal to unity at the
origin. Adjusting the fit to exclude the structure in the central peak does not
significantly change the measured value of the dispersion.

In order to get a better handle on the uncertainties in the results, the north
and south data in the 2dF survey was also divided into two halves and an
identical analysis performed. All results are reported in Table 1.

\begin{figure}[h]
\vbox to5.8in{\rule{0pt}{3.5in}} \includegraphics{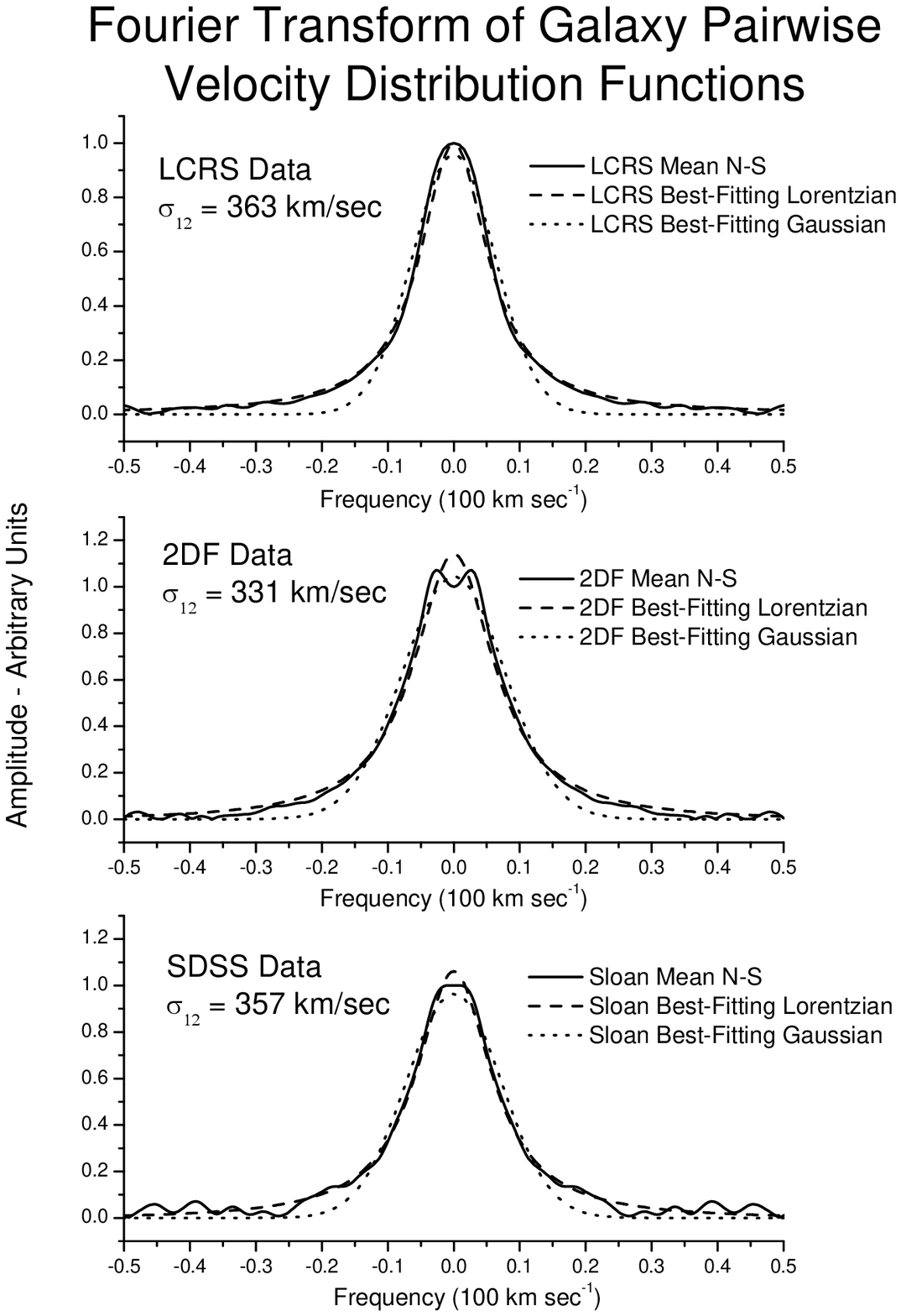} \caption{\small{Figure 1 shows the best
Lorentzian and Gaussian fits to the Fourier transform of the galaxy pairwise
velocity dispersion functions for the LCRS, 2dF, and SDSS Galaxy Redshift
Surveys. In all cases, a Lorentzian is a much better fit to the data. A
Lorentzian is the Fourier transform of an exponential distribution. The value
of the associated exponential is reported.}}
\end{figure}

\section{Discussion}

As is evident in Figure 1, the PVD function for all cases is clearly
well-characterized by an exponential. Furthermore, Table 1 shows that these
measurements are very stable both within and between data sets.

As mentioned in the Introduction and Method sections, one of the major
difficulties in the measurement of the PVD has been the contamination by rich
clusters in the data. For example, Marke \etal (1995) found that by excluding
rich clusters in their data that the signal fell from $\sig= 540\pm180$~\kmsec~
to $\sig=295\pm99$~\kmsec. Additionally two analyses of the LCRS data, one
using a method based on the correlation function and one presented in this {\it
Letter}, give $\sig=452\pm60$~\kmsec (Lin \etal 1995) and $363 \pm 44$~\kmsec~
(LSB98) respectively. It was also shown by Mo \etal (1993) that the estimated
dispersion is extremely sensitive to the presence of rich clusters in a sample,
to the order of errors of several hundred \kmsec for subsets of the same data.

These previous findings emphasize the importance of applying identical methods
to subsets of data and across surveys. The results presented here reflect a
very robust and stable technique. The small variance in the measured signal
across and between data sets clearly indicate that the signal is not being
dominated by rich clusters in the data.

This conclusion is also supported by the agreement between the results of the
two r-band selected surveys, the LCRS and SDSS, both in terms of the value of
the measured signal and the small standard errors. The standard errors of the
2DF b-band selected survey are also of a similar magnitude. Since this latter
b-band survey is expected to contain relatively fewer late-type galaxies, it
would be expected that its standard error would be substantially less than that
of the r-band selected surveys whose signal should be more strongly
contaminated by clusters in the data.

One criticism of this method (see Jing \& B\"{o}rner 1998) is that it does not
adequately take into account infall effects and consequently underestimates the
value of the PVD; compare their value of $570 \pm 80$~\kmsec~ for the LCRS
data. In a complete characterization (see Peebles 1980), the galaxy peculiar
velocity distribution function $f(v_{12}|r)$ is usually modeled as an
exponential along one dimension with

\begin{equation}
f(v_{12}|r)={1\over{\sqrt{2}\sig}} \exp{\left(-{\sqrt{2}|v_{12}-{\overline{
v_{12}}}|\over\sig}\right)}.
\end{equation}

In our analysis, the infall parameter $\overline {v_{12}}$ is not included for
several reasons. Firstly, given the excellent fit of the Fourier transform of
the PVD with the Lorentzian, there is no evidence that the result is being
contaminated by infall effects. Secondly, as was discussed in LSB98, since this
method relies on a Fourier transform of the correlation function and most of
the mass of the correlation function lies within a central core of about 1
\Mpc, the resulting signal will predominately reflect the value of the PVD
within a scale of about 1\Mpc~ where the infall is expected to be small.

In later work, Jing \& B\"{o}rner (2001) re-analyze the LCRS and obtain a
result of $510 \pm 70$~\kmsec. This method very similar to that of LSB98
although the model includes a large-scale infall parameter in the linear regime
characterized by $\beta$. In our method by truncating the correlation function
at approximately 30\Mpc, the approximate scale of the transition between the
linear and non-linear regimes, and windowing the distribution, the signal is
limited to the non-linear regime. This is very important considering that the
distortions of the redshift correlation function due to the small-scale PVD and
the large-scale linear velocity infall are competing effects and so their
errors are positively correlated. Additionally, although Jing \& B\"{o}rner
(2001) incorporate a model for the infall effects of the $\beta$ distortions,
they do include it in the fit but rather set it at a value of $\beta=0.5$.

The competing effects of these two distortions can be clearly seen in Fig. 4 of
Peacock \etal (2001) where a joint maximum likelihood fit of the estimation of
$\sigma_{12}$ and $\beta$ from a larger 2dF data set is presented. The best fit
values are $\sigma_{12} = 385$~\kmsec~and $\beta=0.43\pm0.07$. The LSB98 value
of $363\pm44$~\kmsec~ is within one standard error of this result and would
imply a $\beta$ of $0.4$, while the measurement of Jing \& B\"{o}rner (2001)
$510\pm70$~\kmsec~ implies a best-fitting value of $\beta \simeq 0.6$.

Davis, Miller \& White (1997) also present a method for determining the thermal
velocity dispersion of galaxies using a galaxy-weighted rather than the more
standard pair-weighted measures. An analysis of this result on the LCRS data is
reported in Baker, Davis \& Lin (2000). Although it is difficult to directly
compare their approach to the one presented here, the results are generally
consistent.

\section{Conclusion}

Since galaxy distances and peculiar velocities can not be adequately directly
measured at this point, every approach to measure the properties of the galaxy
peculiar velocity distribution function is somewhat indirect. For example,
almost every method attempts to measure the pairwise galaxy peculiar velocity
distribution function rather the the single particle function itself. Also, it
has been shown that due to the existence of hot clusters in any large redshift
survey, that a direct fit to the correlation function itself is problematic.

The method utilized in this {\it Letter} has several advantages over other
extant techniques. Firstly, it relies on a straightforward Fourier transform of
the $\xi(r_p,\pi)$ correlation function, a mathematical operation that is well
understood. Secondly, in extracting the Fourier transform of the PVD, the need
to model the underlying $\xi(r)$ is obviated. Thirdly, by restricting the
analysis to non-linear scales, the need to model for infall effects is
eliminated. Fourthly, the method returns the Fourier transform of the entire
PVD function and its functional form can be directly investigated. And lastly,
the technique is very stable within and between independent surveys.

Using this method, it has been shown that the galaxy pairwise velocity
distribution function can be well-represented by an exponential for the three
largest existing redshift surveys. The two r-band selected surveys, the LCRS
and the SDSS, give values of with $363 \pm 44$~\kmsec~ and $357 \pm 17$~\kmsec~
respectively. The b-band selected 2dF survey gives slightly lower results with
$331 \pm 19$~\kmsec and this difference is expected given the different
selection effects of r-band and b-band selected surveys. Further analysis of
subsets of the data has shown that the signal is not sensitive to
idiosyncracies in the data sets such as the prevalence of rich clusters.

\acknowledgements

The author would like to acknowledge support from the Jeffress Memorial Trust
and NSF Grant AST 99-00835 and would also like to thank the LCRS, 2dF, and SDSS
surveys for making their data available to the astronomical community. The
author also acknowledges useful discussions with Alexander S. Szalay and thanks
Adrian Pope for random data.

\pagebreak

\begin{deluxetable}{lcccccccc}
\tablewidth{37pc} \tablecaption{Pairwise Peculiar Velocity Dispersion
Measurements (\kmsec)} \tablehead{
        \colhead{Survey}           &
        \colhead{$\sigma_{12}$}   &
        \colhead{$\Delta\sigma_{12}$} &
        \colhead{North} &
        \colhead{South} &
        \colhead{North 1} &
        \colhead{North 2} &
        \colhead{South 1} &
        \colhead{South 2} }
\startdata LCRS & $363$ & $\pm44$ \\  SDSS & $357$ & $\pm17$ & 329 & 353\\ 2dF
& $331$ & $\pm19$ & 314 & 340 & 299 & 328 & 354 & 311 \\ \tablecomments{For all
surveys the mean correlation function of the north and south data was used to
determine the reported dispersion. The standard errors for the LCRS were
calculated from the values for the six separate slices, which are reported in
LSB98. The 2dF and SDSS standard errors were determined from the values for the
north and south subsets. The values for the 2dF data broken up in four subsets
are also shown.}
\enddata
\end{deluxetable}

\end{document}